# Entropy Maximization as a Holistic Design Principle for Complex Optimal Networks and the Emergence of Power Laws

Venkat Venkatasubramanian[1],*, Dimitris N. Politis[2], and Priyan R. Patkar[1]

[1]Laboratory for Intelligent Process Systems, School of Chemical Engineering, Purdue University, West Lafayette, IN 47907, USA

[2]Department of Mathematics, University of California, San Diego, La Jolla, CA 92093, USA

*To whom correspondence should be addressed. Email: venkat@ecn.purdue.edu
Phone: 765-494-0734, Fax: 765-494-0805

**Abstract**

We present a general holistic theory for the organization of complex networks, both human-engineered and naturally-evolved. Introducing concepts of 'value' of interactions and 'satisfaction' as generic network performance measures, we show that the underlying organizing principle is to meet an overall performance target for wide-ranging operating or environmental conditions. This design or survival requirement of reliable performance under uncertainty leads, via the maximum entropy principle, to the emergence of a power law vertex degree distribution. The theory also predicts exponential or Poisson degree distributions depending on network redundancy, thus explaining all three regimes as different manifestations of a common underlying phenomenon within a unified theoretical framework.

## 1. Introduction

Recently, much attention has been paid to the structure of complex networks found in a wide variety of domains such as engineering, biology, economics, ecology, sociology and so on.[1-6] It has been pointed out that these networks, despite the differences in their domains of application or scale, share some common properties. Perhaps the most striking observation is that these networks often display a scale-free or power law behavior of the vertex degree distribution.[1-6] However, the recent studies have largely focused on proposing various mechanisms for the emergence of the power law behavior but not so much on the central question of how the "microscopic" properties of a network such as the vertex degrees of its nodes, which are properties at the individual node level, are related to the "macroscopic" system-level performance measures like network performance or survival. We investigate this question in this paper, and in that process also show another possible mechanism for the power law behavior.

## 2. Modeling Interactions in a Network: Concepts of Value and Satisfaction

Consider a network of interacting members. This may be a computer network such as the Internet, a network of interacting proteins, a social network of friends, an economic network such as a supply chain, and so on. Such a network may be represented as a graph, which is a collection of $N$ nodes or vertices and $E$ links or edges. The nodes represent members of the network and the links between nodes represent interactions or relationships between members. In this paper, we use the terms network and graph, nodes and vertices, interchangeably. The number of edges attached to a vertex is called its degree,[7] $z$. Let $n(z)$ denote the number of vertices with degree $z$ and $p(z)$ the probability that any vertex will have degree $z$. The plot of $n(z)$ vs $z$, or $p(z)$ vs $z$, known as the vertex degree distribution is of great interest, which is an important property as it characterizes the overall topology of the network. As noted for many real-world networks the vertex degree distribution is often found to be a power law, i.e. $p(z)$ is proportional to $z^{-\gamma}$ with the exponent $\gamma$ usually in the range of 2.0–3.0.[1] For random networks it is well known that the degree distribution is Poisson.[7]

In a network, let each member $i$ receive some '*value*' $v_i$ from another member via a relationship or an interaction, as shown in Figure 1. Further, let each member enter into multiple relationships with other partners contributing and deriving values. The notion of value is generic here, which may be, for instance, a design variable like productivity, money, effort, efficiency, time, CPU, bandwidth, emotional support etc., or some combination of such variables. Additionally, there may be various domain-specific constraints on these variables. In general, each such relationship may have associated '*cost*', similar to the expenses associated with creating and maintaining computer lines, roadways, transaction lines, friendships, etc. in real networks. Again, we use cost here in a generic manner.

Let us further introduce the notions of '*intrinsic value*' that a member can contribute in a relationship and '*perceived value*' as felt by the recipient in that relationship, which need not be the same in general. Consider a member *i* building relationships with other members in order to accumulate value for itself from the contributions of its interacting partners. Every member is doing this, but we will examine it from the perspective of member *i*. Now, consider three special cases: (1) the number of partners of *i* is large and the value contributed by any individual partner to *i* is much smaller than the cumulative contribution from all the partners to *i*, (2) the number of partners is quite small and individual contributions form significant portions of the total value accumulated by *I*, and (3) the number of partners is extremely large, an extreme version of Case 1.

**2.1 Large number of partners**
As member *i* begins to build relationships, the first partner *A* it picks up is valued highly (i.e. the *perceived value* is high). We will call the first partner as a *rank one* partner in this relationship and denote it as $A\{(i,1)\}$. The second partner $B\{(i,2)\}$ is valued a little less, the third partner $C\{(i,3)\}$ even less, and the $z^{th}$ partner $X\{(i,z)\}$ considerably less valued compared with the first partner. Thus, the *perceived value* decreases with every incremental contribution. This is a reasonable assumption as such diminishing returns are encountered in many practical applications when a large number of entities contribute small amounts each. Let $v_m$ denote the perceived value felt by member *i* from its partner of rank *m*. We have not explicitly considered the cost of relationships here but handled it implicitly, in the sense, the value measure may be treated as a *net value* after accounting for cost. Although we motivate the notion of rank as the sequential order of a partner in a set of relationships, it need not be restricted to this role alone. In general, the rank of a partner is an indicator of its relative importance in comparison with other partners in a set of relationships.

We call the cumulative sum of all the perceived values the '*local satisfaction*' measure *S*, which can be thought of as a local performance metric. We postulate this satisfaction measure to be finite as the number of relationships tends to infinity. This is again a reasonable requirement as most real-life situations involve some kind of a saturation effect in such interactions and the notion of diminishing returns, as noted earlier, is a commonly used principle in many applications. Thus we have:

$$S_\infty = v_1 + v_2 + v_3 + ... = \sum_{m=1}^{\infty} v_m < \infty \qquad (1)$$

For the sake of simplicity, let all members contribute the same *intrinsic value v*. In a large population of similar members, this is again a reasonable assumption. Thus, everyone contributes the same intrinsic value *v*, but it is perceived differently by others depending on how many other partners they already have, i.e. depending on the rank of the relationship. As mentioned earlier, the perceived value of a partner decreases with increasing rank in that particular relationship, which can be modeled as follows:

$$v_m = v\left[\frac{1}{m^q}\right] \quad \text{where } m \text{ is the rank and } v_m \text{ is the perceived value} \tag{2}$$

Note that the same partner $X$ may be valued differently in different relationships depending on its rank in those relationships, even though $X$ contributes exactly the same intrinsic value $v$ in all those relationships. Therefore, the satisfaction $S$ for member $i$ from all its $z$ contributing relationships (partners) is given by:

$$S_z = v \sum_{m=1}^{z} \frac{1}{m^q} \tag{3}$$

Recall that in the limit $z \to \infty$, we require this sum to converge to a finite value:

$$S_\infty = v \sum_{m=1}^{\infty} \frac{1}{m^q} < \infty \tag{4}$$

This implies that the exponent $q > 1$, as the series will otherwise diverge. Although $q$ can take on any number of values satisfying this requirement, all those values place additional demands or expectations on the behavior of the function $S$. The *least restrictive* requirement corresponds to the limiting case $q = 1 + \varepsilon$, where $\varepsilon > 0$ but arbitrarily small. Under this requirement, we have:

$$S_\infty = v \sum_{m=1}^{\infty} \frac{1}{m^{1+\varepsilon}} < \infty \tag{5}$$

As $\varepsilon \to 0$, for large but finite $z$, the resulting harmonic series is approximated by:

$$S_z = v \sum_{m=1}^{z} \frac{1}{m} \approx v[\ln z + 0.577] \tag{6}$$

where 0.577 is known as the Euler's constant. Setting $\varepsilon \to 0$ may be considered as the *asymptotic limit* of the saturation function $S$. Since $v$ is an arbitrary quantity, we set $v=1$ to simplify the analysis. Thus, in the limit $\varepsilon \to 0$ the cumulative perceived value, i.e. the local satisfaction measure $S_z$ for any member, increases approximately as $\ln z$ for large $z$.

So far we have only considered one member, namely, $i$. However, every member $j$ in this network tries to accumulate satisfaction given by Eqn. 6; therefore, in general (letting $v=1$):

$$S_z^{(j)} = \sum_{m=1}^{z^{(j)}} \frac{1}{m} \approx \ln z^{(j)} + 0.577 \tag{7}$$

where $z^{(j)}$ is the number of partners of member $j$. Every member tries to increase its local satisfaction to the extent possible, subject to cost and other constraints. Thus, there will be a distribution of local satisfaction measures for the entire network.

From a holistic design perspective, both human-engineered and naturally evolved networks are organized to perform certain overall function(s) well, i.e. designed to meet certain performance criteria or survival objectives as a whole. Thus, one is not overly concerned with the local satisfaction or performance of any particular member but with the overall satisfaction or performance level in the entire network. A simple intuitive measure of this overall satisfaction level is the average of all the local satisfaction measures. We call this average the *global satisfaction measure*, $S_G$, given by

$$S_G = \langle S^{(j)} \rangle = 0.577 + \langle \ln z^{(j)} \rangle \tag{8}$$

The global satisfaction measure may be thought of as a performance metric of the entire network. Thus, from a holistic perspective, the design principle for the organization of the network would be to meet a certain level of global satisfaction or performance $\theta^G$ as the design target:

$$S_G = \theta^G = 0.577 + \langle \ln z \rangle = 0.577 + \theta \quad \text{where } \theta = \langle \ln z \rangle \tag{9}$$

Therefore, the organization of the network, i.e. the distribution of links in the network will be constrained by the design criterion in Eqn. 9.

There is an additional design criterion; one would like the network to be designed such that it can survive or meet its performance target under a wide variety of environments or operating conditions. For example, computer or communication networks would be designed to function and meet performance target criteria under a variety of anticipated operating conditions involving different communication situations, demand patterns, failures or additions of nodes and edges, etc. A reliable design would not be limited to perform well only under certain conditions, i.e. one would not want to *bias* the design for a specific operating environment, particularly if the nature of future environments is unknown, uncertain or unpredictable. Hence, the network design should reflect this *inherent uncertainty* about future operating environments and *minimize the bias* or any unwarranted assumptions about them. In other words, a reliable design should accommodate as much uncertainty about the future operating environments as allowed by the constraints. Thus, we have two "macroscopic" system-level design criteria determining the structure of a network: the global satisfaction measure target and reliable performance under maximum uncertainty. Combining these two criteria, the network design problem can now be posed as the following question: *What is the least biased distribution of links in a network such that the constraint $S_G = \theta^G$, or equivalently $\langle \ln z \rangle = \theta$, is satisfied?*

## 2.2 Entropy Maximization as a Holistic Design Principle:

The above question can be answered by applying the Maximum Entropy Principle,[8-10] which states that given some partial information about a random variate, of all the distributions consistent with the given information or constraints, the *least biased distribution* is the one having the *maximum entropy* associated with it. In the context of the discussion above, maximizing entropy is the same as maximizing uncertainty in the information theoretic sense.[11] Applying the maximum entropy formulation to Eqn. 9 the least biased distribution satisfying the constraint $\langle \ln z \rangle = \theta$ is a power law given by:

$$\Pr(Z = z) \approx \frac{1}{\theta} \cdot \frac{1}{(z)^\gamma} \quad \text{where } \gamma = \frac{1+\theta}{\theta} \text{ for large } z \tag{10}$$

We will presently derive this result. Consider a continuous random variate $Z$. Let $z \geq 1$ and $y = \ln z \geq 0$. Under the constraint $\langle y \rangle = \theta$, for some positive real number $\theta$, it is well known that the maximum entropy formulation yields the exponential probability density function and the cumulative distribution function respectively given below: [8,9]

$$f_Y(y) = \frac{1}{\theta} e^{-\left(\frac{y}{\theta}\right)} \quad \text{and} \quad F_Y(y) = 1 - e^{-\left(\frac{y}{\theta}\right)} \quad \text{for } y \geq 0 \tag{11}$$

But $F_Z(z) = \Pr(Z \leq z) = \Pr(\ln Z \leq \ln z) = \Pr(Y \leq y) = 1 - e^{-\left(\frac{y}{\theta}\right)} = 1 - e^{-\left(\frac{\ln z}{\theta}\right)} = 1 - z^{-\left(\frac{1}{\theta}\right)}$

Therefore, $f_Z(z) = \frac{1}{\theta} \cdot \frac{1}{z^{\left(\frac{1+\theta}{\theta}\right)}} = \frac{1}{\theta} \cdot z^{-\gamma}$, where $\gamma = \frac{1+\theta}{\theta}$

Using this power law density, we can also easily show that

$$\gamma = \frac{2\langle z \rangle - 1}{\langle z \rangle - 1}, \quad \text{where } \langle z \rangle = \int_1^\infty z f_Z(z) dz \tag{12}$$

Finally, in the practical case where $Z$ is taking values $z$ in the positive integers, the power law probability density of Eqn. 11 can be discretized with the help of Taylor's theorem to yield the discrete power law:

$$\Pr(Z = z) \approx \frac{1}{\theta} \cdot \frac{1}{(z)^\gamma}, \quad \text{where } \gamma = \frac{1+\theta}{\theta} \text{ for large } z \tag{13}$$

The importance of Eqn. 12 lies in the fact that the mean value of $Z$ is a quantity easily estimated in practice using observed sample averages. To estimate $\theta$ from data, the constraint $\langle \ln z \rangle = \theta$ may directly be utilized. In other words, $\theta$ may be estimated as the sample average of ln $z$ from the data. However, if the sample average of ln $z$ is unavailable but the sample average of $z$ is, then Eqn. 12 may be used to estimate $\theta$ and $\gamma$.

Nevertheless, it should be noted that if and when the true value of $\theta$ is close to one (and the true value of $\gamma$ is close to 2), the sample average of $z$ can be an unreliable estimator of its theoretical expectation since the latter can be huge in that case. It is easy to see from Eqn. 12 that the bounds on $\gamma$ for a connected network are $2 < \gamma < 3.02$.

Thus, when the network design criterion is to meet a target global satisfaction or performance measure for a wide variety of operating environments, in the asymptotic limit for large networks, power law emerges as the *optimal* distribution of network connections. It is optimal in the sense that this distribution allows for maximum uncertainty about future operational environments, and therefore, optimally equips the network to handle various possible future scenarios. Since the power law emerges under some very general conditions, without any domain specific details or constraints playing a role, it appears to be a universal behavior explaining its prevalence in a wide variety of complex systems.

Since Eqn. 12 estimates $\gamma$ in terms of $\langle z \rangle$, we compare this estimation ($\gamma_{est}$) with results reported in the literature[1] for several networks, as shown in Table 1. The published data[1] does not separately provide $\langle z_{in} \rangle$ and $\langle z_{out} \rangle$ but only the overall $\langle z \rangle$; therefore, $\gamma_{est}$ is also such an overall value. The subscripts 'in' and 'out' respectively refer to directed links coming in to and going out of a node. Since $\langle z \rangle$ is independent of $N$ in Eqn. 12, the estimation of $\gamma$ is valid for all network sizes provided they satisfy the general conditions discussed earlier.

We are not surprised by the disagreement with the $\gamma$ values for food webs (Ythan Estuary and Silwood Park). We believe that food webs belong to a different class of complex networks because of the special predator-prey relationship between nodes. For the networks considered in the analysis, it is always better for a member (node) to have more relationships (links) as they all contribute value, however small. But in food webs, each species may like to have more prey but fewer predators. This conflict makes the situation asymmetric. Hence, we believe the analysis does not directly apply to food webs; however, we have adapted the same to model food webs, which will be presented in a forthcoming publication.

**2.3 Small number of partners**
We have so far considered networks where the number of links (partners) per node or the link density $E/N$, is large. It is instructive to explore the theory's predictions in two other limiting cases, one being a small number of links per node. Since links generally have an associated cost, such a situation is typically encountered under very high cost per link, where practical resource limitations limits the total number of links to a small number for a given $N$. Each member then has, on an average, only a small number of relationships and, therefore, in contrast to Case 1, perceives every relationship as more or less equally valuable. Under this condition, far below the saturation limit, the log function $\ln z$ may be approximated linearly and hence the cumulative local satisfaction measure represented as:

$$S_{small} \approx \upsilon z \tag{14}$$

Note, Eqn. 14 will be valid even when the number of relationships per node is relatively high, provided the satisfaction function is far below the saturation limit. Thus, the key requirement here is the satisfaction measure growing at least approximately linearly with additional relationships, and the diminishing returns regime as modeled by the log function not having set in for any nodes in the network. Therefore, the global satisfaction measure is given by (for $v=1$):

$$S_{G\,small} = \langle S_{small} \rangle = \langle z \rangle = \theta_{small} \tag{15}$$

Under this constraint, the Maximum Entropy Principle yields the *exponential distribution* as the least biased distribution given by:

$$f_Z(z) = \frac{1}{\theta_{small}} e^{-\frac{z}{\theta_{small}}} \quad \text{and} \quad F_Z(z) = 1 - e^{-\frac{z}{\theta_{small}}}, \; z \geq 0 \tag{16}$$

**2.4 Extremely large number of partners**

In the other limit of extremely large number of links per node, we expect the satisfaction measure to saturate much faster than the log function approximation. Thus, after all members develop some $z_{sat}$ number of relationships, additional relationships (links) bring *zero* value and may, therefore, be formed randomly between any two nodes with equal probability leading to a Poisson distribution. Thus, the theory predicts that three distinct classes of networks should arise for three different functional behaviors of the local satisfaction measure. Exponential networks correspond to the class that emerges when the satisfaction measure grows approximately linearly with additional relationships, power law networks when the dependence is log-like, and Poisson networks when saturation sets in. These three classes can also appear as three regimes in a given network as the network moves from a sparely connected state to reasonably redundant to extremely redundant in the number of links. In such a situation, the exponential regime has a relatively narrow range due to the limited validity of the linearity approximation of the log function in Eqn. 14. It is, however, more difficult to determine the upper bound for the power law regime *a priori* since the rate at which the satisfaction measure saturates may vary from network to network.

The existence of these three regimes as a function of redundancy was recently pointed out by Venkatasubramanian *et al.*[12] but without an explanation for their occurrence. Furthermore, exponential degree distributions have been reported for real networks when the cost constraints of adding new links are particularly severe as noted by the authors in [13]. The disappearance of power law is also known to occur at the other end of the spectrum when the network is extremely redundant in number of links;

Barabási *et al.*[3] found that starting with *N* nodes and no links, and adding links incrementally, the power-law scaling was found at early times but was not stationary and eventually disappeared. Since most real-world networks are designed to be reasonably redundant to make them more robust and efficient, they all typically fall in the middle regime and thus the apparent ubiquitous occurrence of power laws and their seeming universality.

**3. Summary**

In conclusion, we have presented a general holistic theory for the organization of complex networks that assumes that complex networks, both human-engineered and naturally evolved, are organized to meet certain design or survival objective(s) for a wide variety of operating or environmental conditions. For a large class of networks, where the interactions between members are of mutual benefit or value, the design objective is to achieve a desired overall network performance as modeled by an average cumulative interaction value. Under some general conditions, we have shown that for large networks in the asymptotic limit of local performance saturation, the design requirement of reliable performance under maximum uncertainty leads to the emergence of power laws as a consequence of the maximum entropy principle. That is, under these general conditions, a power law-based organization gives a network the *maximum flexibility* to perform well overall in a wide variety of operating environments. Note that for a specified operating environment, there may exist some other distribution that can outperform the maximum entropy distribution with respect to the global performance target; however, such a biased network may fail when the underlying environment changes, whereas the maximum entropy distribution-based network will continue to survive and perform. Thus, under entropy maximization, the network's performance is optimized to accommodate a wide variety of future environments whose nature is unknown, unknowable and hence uncertain.

In this context, it is interesting to consider the implications of this result for a well known economic phenomenon, namely, the Pareto distribution of wealth. It has been known for nearly a century that the distribution of wealth in many countries follows a power law with $2 < \gamma < 3$ as first observed by the Italian engineer Vilfredo Pareto.[14] The main consequence of this distribution is the inequity in the allocation of wealth, i.e. a small number of people hold the majority of the wealth in a society. While this may seem unfair, our theory suggests that this is the optimal distribution of wealth for the society at large. Given the design criterion that a society needs to remain economically productive meeting some global productivity or performance targets for its survival under a wide variety of future economic environments whose nature is unknown, unpredictable and hence uncertain, the Pareto distribution is the optimal way of distributing wealth in the face of an uncertain future. Note that this power law result also implies that the local satisfaction measure, which here is a function of a given individual's (node's) accumulated wealth (the proxy for partners), varies as a log function of wealth. While the log function is not typically used to model wealth effects in economics as it does not saturate in the infinite limit, the point here is that the approximate saturation-like behavior displayed by the log function for large, but not infinite, amount of wealth is perhaps a reasonable enough approximation in modeling this economic phenomenon that the theory's prediction of a power law as the optimal distribution is observed in reality as the Pareto distribution.

The generality of this theory explains the apparent universality of power laws in many complex systems. The theory also predicts the emergence of exponential and Poisson regimes as a function of the redundancy of a network, thus explaining all three regimes as different manifestations of the same underlying phenomenon within a unified theoretical framework.

The main contribution of this paper is to present a theoretical framework to address one of the most important questions in complex networks organization, namely, how the "microscopic" properties of a network such as the vertex degree are determined by, and conversely determine, the "macroscopic" or system-level properties such as network performance. This is an important question to resolve as it can pave the way for an integrative perspective on complex networks and systems by relating node or subsystem-level properties to system-level ones, which has significant implications in a variety of domains such as biology, engineering, economics, sociology, etc. This theory is an attempt to discover the equivalent of the partition function in statistical mechanics, which relates the molecular level properties of a gas to the macroscopic thermodynamic properties. Thus, it has not escaped our notice that the functional form of the satisfaction measure resembles that of entropy and information, and that the maximum entropy principle plays such a central role. It appears that the principle that governs the organization of inanimate physical systems is also at play for the organization of animate biological systems.

**Acknowledgements:**

The authors would like to thank Dr. Santhoji Katare, Prof. Suresh Jagannathan and Prof. Ananth Iyer for helpful discussions.

Table 1. A comparison of the estimated and observed γ values for various networks.

| Network | Size | $\langle z \rangle$ | $\gamma_{out}$ | $\gamma_{in}$ | $\gamma_{est}$ |
|---|---|---|---|---|---|
| WWW | 325,729 | 4.51 | 2.45 | 2.1 | 2.28 |
| WWW | $4 \times 10^7$ | 7 | 2.38 | 2.1 | 2.17 |
| WWW | $2 \times 10^8$ | 7.5 | 2.72 | 2.1 | 2.15 |
| Internet, domain | 3,015 – 4,389 | 3.42 – 3.76 | 2.1 – 2.2 | 2.1 – 2.2 | 2.36 – 2.41 |
| Internet, router | 3,888 | 2.57 | 2.48 | 2.48 | 2.63 |
| Internet, router | 150,000 | 2.66 | 2.4 | 2.4 | 2.60 |
| Movie actors | 212,250 | 28.78 | 2.3 | 2.3 | 2.04 |
| Coauthors, SPIRES | 56,627 | 173 | 1.2 | 1.2 | 2.00 |
| Coauthors, neuro. | 209,293 | 11.54 | 2.1 | 2.1 | 2.09 |
| Coauthors, math. | 70,975 | 3.9 | 2.5 | 2.5 | 2.34 |
| Metabolic, E. coli | 778 | 7.4 | 2.2 | 2.2 | 2.16 |
| Protein, S. cerev. | 1870 | 2.39 | 2.4 | 2.4 | 2.72 |
| Ythan Estuary | 134 | 8.7 | 1.05 | 1.05 | 2.13 |
| Silwood Park | 154 | 4.75 | 1.13 | 1.13 | 2.27 |
| Citation | 783,339 | 8.57 | | 3 | 2.13 |
| Phone-call | $53 \times 10^6$ | 3.16 | 2.1 | 2.1 | 2.46 |
| Words, co-occurrence | 460,902 | 70.13 | 2.7 | 2.7 | 2.01 |
| Words, synonyms | 22,311 | 13.48 | 2.8 | 2.8 | 2.08 |

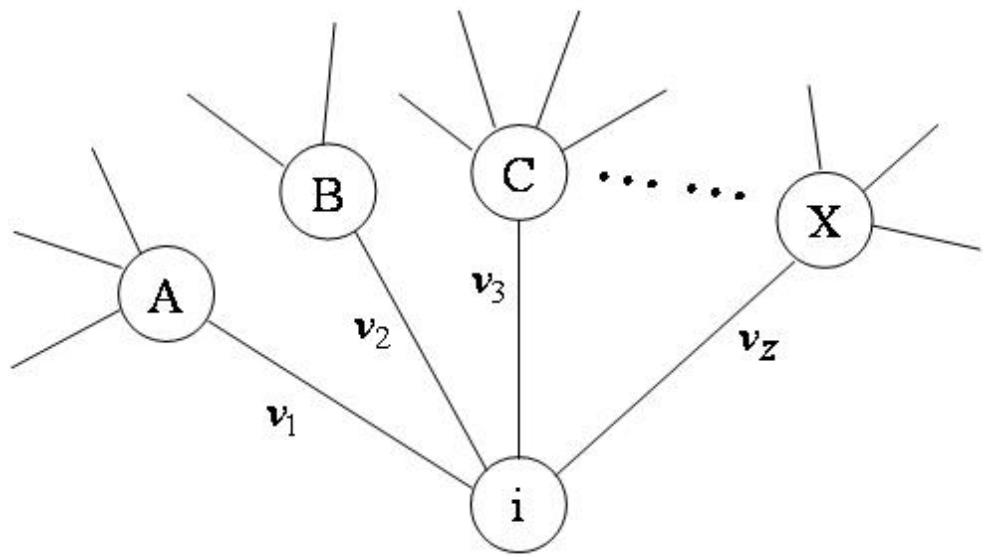

Fig. 1. Perceived values of interaction at an arbitrary node *i*.